\documentclass[aps,prd,preprint,superscriptaddress,showpacs]{revtex4}
\usepackage{graphicx}

\usepackage{ulem}
\usepackage{color}
\definecolor{My_red}        {cmyk}{0.00,1.00,1.00,0.20}

\newcommand{\bmat}{\left(\begin{array}}
\newcommand{\emat}{\end{array}\right)}
\newcommand{\beq}{\begin{equation}}
\newcommand{\eeq}{\end{equation}}

\def\nnb{\nonumber}

\def\bwt{\begin{widetext}}
\def\ewt{\end{widetext}}
\def\be{\begin{equation}}
\def\ee{\end{equation}}
\def\bea{\begin{eqnarray}}
\def\eea{\end{eqnarray}}
\def\bean{\begin{eqnarray*}}
\def\eean{\end{eqnarray*}}
\def\bary{\begin{array}}
\def\eary{\end{array}}
\def\bit{\begin{itemize}}
\def\eit{\end{itemize}}

\def\su5u1{SU(5) \times U(1)}
\def\fsu5u1{SU(5) \times U(1)'}
\def\so10{SO(10)}
\def\sq20{SO(10) \times SO(10)}

\def\bwt{\begin{widetext}}
\def\ewt{\end{widetext}}
\def\be{\begin{equation}}
\def\ee{\end{equation}}
\def\bea{\begin{eqnarray}}
\def\eea{\end{eqnarray}}
\def\bean{\begin{eqnarray*}}
\def\eean{\end{eqnarray*}}
\def\bary{\begin{array}}
\def\eary{\end{array}}
\def\bit{\begin{itemize}}
\def\eit{\end{itemize}}

\def\su5u1{SU(5) \times U(1)}
\def\fsu5u1{SU(5) \times U(1)'}
\def\so10{SO(10)}
\def\sq20{SO(10) \times SO(10)}

\usepackage[centertags]{amsmath}
\usepackage{amssymb}

\begin{document}

\title{ $SU(5)$ and $SO(10)$ Models from F-Theory with Natural Yukawa Couplings}

\author{Tianjun Li}

\affiliation{Key Laboratory of Frontiers in Theoretical Physics, 
      Institute of Theoretical Physics, Chinese Academy of Sciences, 
Beijing 100190, P. R. China }

\affiliation{George P. and Cynthia W. Mitchell Institute for
Fundamental Physics, Texas A$\&$M University, College Station, TX
77843, USA }

\date{\today}

\begin{abstract}

We construct the $SU(5)$ and $SO(10)$ models from F-theory. 
Turning on the $U(1)$ fluxes, we can break the $SU(5)$ 
gauge symmetry down to the Standard Model (SM) gauge 
symmetry, and break the $SO(10)$ gauge symmetry down to 
the $SU(3)_C\times SU(2)_L\times SU(2)_R \times U(1)_{B-L}$
gauge symmetry. In particular, all the SM fermion
Yukawa couplings preserve the enhanced
$U(1)_a \times U(1)_b$ gauge or global symmetries 
at the triple intersections of the SM fermion and 
Higgs curves. And the SM fermion masses
and mixings can be generated in the presence of 
background fluxes. In our models, the doublet-triplet 
splitting problem can be solved naturally.
The additional vector-like 
particles can obtain heavy masses via the instanton effects
or Higgs mechanism and then decouple at the high scale. 
The SM gauge couplings at the string scale, which are 
splitted due to the $U(1)$ flux effects, can be explained 
by considering heavy threshold corrections 
from the extra vector-like particles. Moreover, in the 
$SU(5)$ model, we have the Yukawa coupling unification 
for the bottom quark and tau lepton. In the $SO(10)$ models, we 
have the Yukawa coupling unification for the top and 
bottom quarks, and the Yukawa coupling unification 
for the tau lepton and tau neutrino.

\end{abstract}

\pacs{11.10.Kk, 11.25.Mj, 11.25.-w, 12.60.Jv}

\preprint{MIFP-09-24}

\maketitle

\section{Introduction}

The great challenge in string phenomenology is to construct 
realistic string models with moduli stabilization and 
without extra chiral exotics, and then make clean 
predictions that can tested at the Large Hadron Collider 
(LHC) and other future experiments.  Previously, 
string model building has been studied extensively in
 the heterotic $E_8\times E_8$ string theory and 
Type II string theories with D-branes.

Recently, semi-realistic Grand Unified Theories (GUTs) have been 
constructed locally in 
the F-theory with seven-branes, which can be considered as the
strongly coupled formulation of ten-dimensional Type IIB string 
theory with a varying axion ($a$)-dilaton ($\phi$) field 
$\tau=a+ie^{-\phi}$~\cite{Vafa:1996xn, Donagi:2008ca,
Beasley:2008dc, Beasley:2008kw, Donagi:2008kj}
(For a briefly review, see Section II.). Then further model
building and phenomenological consequences have been studied 
extensively~\cite{Heckman:2008es, Marsano:2008jq, Marsano:2008py,
Heckman:2008qt, Font:2008id, Braun:2008pz, Heckman:2008qa,
Jiang:2008yf, Collinucci:2008zs, Blumenhagen:2008aw, Heckman:2008jy,
Bourjaily:2009vf, Hayashi:2009ge, Andreas:2009uf, Chen:2009me,
Heckman:2009bi, Donagi:2009ra, Bouchard:2009bu, Randall:2009dw,
Heckman:2009de, Marsano:2009ym, Bourjaily:2009ci, 
Tatar:2009jk, Jiang:2009za, Marsano:2009gv, Cecotti:2009zf}. 
As we know, the GUTs without additional chiral 
exotic particles are asymptotically free, and asymptotic 
freedom can be translated into the existence of a consistent 
decompactification limit. Moreover, the Planck scale $M_{\rm Pl}$ 
is about $10^{19}$ GeV while the GUT scale $M_{\rm GUT}$
is around $2\times 10^{16}$ GeV, so $M_{\rm GUT}/M_{\rm Pl}$ is 
indeed a small number around $10^{-3}-10^{-2}$.
Thus, from the effective field theory point of view in 
the bottom-up approach, it is natural to assume that
$M_{\rm GUT}/M_{\rm Pl}$ is small, and
then gravity can be decoupled.
In the decoupling limit where $M_{\rm Pl} \rightarrow \infty$
while $M_{\rm GUT}$ remains finite, semi-realistic
$SU(5)$ models and $SO(10)$ models without chiral exotic particles
have been constructed locally. To decouple gravity and 
avoid the bulk matter fields on the observable 
seven-branes, we can show that the observable seven-branes should 
wrap a del Pezzo $n$ surface $dP_n$ with $n \ge 2$ for the internal 
space dimensions (For a review of del Pezzo $n$ surfaces, 
see Appendix A.)~\cite{Beasley:2008dc, Beasley:2008kw}. 
The GUT gauge fields are on the worldvolume of the
observable seven-branes, while the matter and 
Higgs fields are localized on the codimension-one curves 
in $dP_n$. 

A brand new feature is that the $SU(5)$ gauge symmetry
can be broken down to the Standard Model (SM) gauge symmetry
by turning on $U(1)_Y$ flux~\cite{Beasley:2008dc, Beasley:2008kw}, 
and the $SO(10)$  gauge 
symmetry can be broken down to the $SU(5)\times U(1)_X$
and $SU(3)\times SU(2)_L\times SU(2)_R\times U(1)_{B-L}$
gauge symmetries by turning on the $U(1)_X$ and $U(1)_{B-L}$
fluxes, respectively~\cite{Beasley:2008dc, Beasley:2008kw,
Jiang:2008yf, Jiang:2009za, Font:2008id}. 
In the $SU(5)$ models, 
to generate the up-type quark Yukawa couplings 
$\mathbf{10}_i \mathbf{10}_j \mathbf{5}$ at the
triple intersections where the
gauge symmetry is enhanced to $E_6$, the matter curves
for $\mathbf{10}_i$ are pinched~\cite{Beasley:2008dc, Beasley:2008kw}.  
To be concrete, the
two matter curves for $\mathbf{10}_i$ and $\mathbf{10}_j$
that emanate from an $E_6$ point are acturally two different
parts of a single curve which is pinched at the $E_6$ point.
In other words, it can be viewed as starting with two
distinct cuves which connect to each other some distance
away from the $E_6$ point. This connection is not nicely
captured by the local field theory description of the $E_6$
point in terms of the breaking 
$E_6 \rightarrow SU(5)\times U(1)_a\times U(1)_b$ as only a certain
linear combintaion of $U(1)_a$ and $ U(1)_b$ may be realized as actual
gauge symmetry. Simiar results hold for the $SO(10)$ models
where the gauge symmetry is broken down to the flipped 
$SU(5)\times U(1)_X$ gauge symmetry by turning on the $U(1)_X$ 
flux~\cite{Beasley:2008dc, Beasley:2008kw,
Jiang:2008yf, Jiang:2009za}.
Although the SM fermion Yukawa martices are rank one, 
the SM fermion masses and mixings
can be generated in the presence of 
$H$-fluxes~\cite{Heckman:2008qa, Cecotti:2009zf}. 
However, to construct the globally consistent models, one
found that the problematic situation with two different
$\mathbf{10}_i$ and $\mathbf{10}_j$ matter curves meeting at 
an $E_6$ point requires significant fine-tuning to achieve
at all. What happens most generically in globally consistent
models is that an $E_6$ point occurs when a single $\mathbf{10}$
matter curve meets a single $\mathbf{5}$ Higgs curve.
Of the two $U(1)_a\times U(1)_b$ symmetries that one might
expect in the small open neighborhood of the $E_6$ point,
one linear combination of them is essentially projected out
by a monodromy action. Because $\mathbf{10}_i$ and
 $\mathbf{10}_j$ carry the same charge under the 
other linear combination of $U(1)_a\times U(1)_b$ symmetries
that is not broken
in the small open neighborhood of the $E_6$ point, 
the up-type quark Yukawa couplings
can be realized~\cite{Hayashi:2009ge, Donagi:2009ra, 
 Marsano:2009gv}. Moreover, in the $SU(5)$ models,
 the SM gauge couplings at the string scale are splitted
due to the $U(1)_Y$ flux~\cite{Donagi:2008kj, Blumenhagen:2008aw}, 
thus, it is pretty difficult to explain 
this splitting in the minimal $SU(5)$ models without 
extra vector-like particles.
In the $SO(10)$ models where the gauge symmetry is broken
down to the $SU(3)\times SU(2)_L\times SU(2)_R\times U(1)_{B-L}$
gauge symmetry by turning on the $U(1)_{B-L}$ flux~\cite{Font:2008id},
three right-handed lepton doublets, two left-handed
quark doublets (the first two generations) and one right-handed 
quark doublet (the third generation) are on one matter
curve, while three left-handed lepton doublets, two  
right-handed quark doublets (the first two generations), 
and one left-handed quark doublet (the third generation)
are on the other matter curve. Although the correspoding
$U(1)_a\times U(1)_b$ gauge or global 
symmetries at the triple intersections of the SM fermion and 
Higgs curves can be preserved explicitly, one indeed needs some
mechanism to generate the small CKM quark
mixings between the first two generations and the third
generation.

In this paper, we first briefly review the F-theory model 
building. We construct one $SU(5)$ model and two $SO(10)$ 
models from F-theory. In our models, we seek to retain 
a number of $U(1)$ symmetries from the underlying $E_8$
structure while ensuring that all the possible SM fermion
Yukawa couplings ({\it i.e.}, all the entries in all 
the SM fermion Yukawa matrices) are allowed to be non-zero
by the corresponding selection rules. This is desirable
from the low energy phenomenological point of view. 
In our models, we can break the $SU(5)$ gauge 
symmetry down to the SM gauge symmetry by turning on 
the $U(1)_Y$ flux, and break the $SO(10)$ gauge symmetry 
down to the 
$SU(3)_C\times SU(2)_L\times SU(2)_R\times U(1)_{B-L}$
gauge symmetry by turning on the $U(1)_{B-L}$ flux. 
To preserve the explicit
$U(1)_a\times U(1)_b$ gauge or global 
symmetries at the triple intersections of the SM fermion and 
Higgs curves, we put the left-handed quarks and
the right-handed quarks on the different matter curves,
and put the left-handed leptons and neutrinos and
the right-handed leptons and neutrinos on the
different matter curves. Thus, in our models,
all the SM fermion Yukawa couplings 
are invariant under the enhanced $U(1)_a \times U(1)_b$ 
gauge or global symmetries at the intersections of 
the SM fermion and Higgs curves. And then our models
can be understood very well from the local field theory
description. Although the SM fermion Yukawa matrices
are rank one, the SM fermion
masses and mixings can be generated in the presence
of $H$-fluxes~\cite{Cecotti:2009zf}.
Moreover, we can solve the doublet-triplet splitting
problem naturally.  The extra vector-like 
particles can obtain the heavy masses via the instanton effects
or Higgs mechanism and then decouple at
the high scale. Therefore,
we only have the supersymmetric Standard Model 
at the low energy.
Similar to the $SU(5)$ 
models~\cite{Donagi:2008kj, Blumenhagen:2008aw},
we show that the SM gauge couplings 
at the string scale are also splitted
in our $SO(10)$ models due to the $U(1)_{B-L}$ flux.
Interestingly, the extra vector-like particles in
our $SU(5)$ and $SO(10)$ models can be considered as
the heavy threshold corrections to the renormalization
group equation (RGE) running for the SM gauge couplings,
and then we can explain 
the SM gauge couplings at the string scale elegantly.
Furthermore, in the $SU(5)$ model, 
we can have the Yukawa coupling unification for the bottom 
quark and tau lepton. In the $SO(10)$ models, we can
have the Yukawa coupling unification for the top and 
bottom quarks, and the Yukawa coupling unification 
for the tau lepton and tau neutrino.

This paper is organized as follows. In Section II, we briefly
 review the F-theory model building. In Sections III and IV, 
we construct the $SU(5)$ model and the $SO(10)$ models, respectively.
Our conclusions are in Section V.
In Appendix A, we briefly review the del Pezzo surfaces.





\section{F-Theory Model Building}

We first briefly review the F-theory model 
building~\cite{Vafa:1996xn, Donagi:2008ca,
Beasley:2008dc, Beasley:2008kw, Donagi:2008kj}.
The twelve-dimensional F theory is a convenient way to describe 
Type IIB vacua with varying
axion-dilaton $\tau=a+ie^{-\phi}$. We compactify F-theory on a
Calabi-Yau fourfold, which is elliptically fibered $\pi: Y_4 \to B_3$
with a section $\sigma: B_3 \to Y_4$. The base $B_3$ is the internal
space dimensions in Type IIB string theory, and the complex structure
of the $T^2$ fibre encodes $\tau$ at each point of $B_3$. The SM or GUT
gauge theories are on the worldvolume of the observable
seven-branes that wrap
a complex codimension-one suface in $B_3$. Denoting the complex
coordinate tranverse to these seven-branes in $B_3$ as $z$, we can
write the elliptic fibration in  Weierstrass form
\begin{eqnarray}
 y^2=x^3+f(z)x+g(z)~,~\,
\end{eqnarray}
where $f(z)$  and $g(z)$ are sections of $K_{B_3}^{-4}$ and
$K_{B_3}^{-6}$, respectively. The complex structure of the fibre is 
\begin{eqnarray}
 j(\tau)~=~ {{4(24f)^3}\over {\Delta}}~,~~~
\Delta~=~ 4 f^3 + 27 g^2 ~.~\,
\end{eqnarray}
At the discriminant locus $\{\Delta=0\} \subset B_3$,
the torus $T^2$ degenerates by pinching one of its
cycles and becomes singular. For a generic pinching one-cycle 
 $(p, q)=p\alpha+q\beta$ where $\alpha$ and $\beta$
are one-cylces for the torus $T^2$, we obtain a $(p,q)$ seven-brane in 
the locus where the $(p,q)$ string can end.
The singularity types of the ellitically fibres fall into the 
familiar $ADE$ classifications, and we identify the corresponding 
$ADE$ gauge groups on the seven-brane world-volume. 
This is one of the most important advantages for the F-theory model building: 
the exceptional gauge groups appear rather naturally, which is absent in
the perturbative Type II string theory. And then all the SM fermion Yuakwa
couplings in the GUTs can be generated.

We assume that the observable seven-branes with GUT models
on its worldvolume wrap a complex codimension-one 
suface $S$ in $B_3$, and the observable gauge symmetry
is $G_S$. When $h^{1,0}(S)\not=0$, the low energy
spectrum may contain the extra states obtained
by reduction of the bulk supergravity modes of
compactification. So we require that 
$\pi_1(S)$ be a finite group. In order to decouple
gravity and construct models locally, the extension 
of the local metric on $S$ to
a local Calabi-Yau fourfold must have a limit where
the surface $S$ can be shrunk to zero size. This implies
that the anti-canonical bundle on $S$ must be ample. 
Therefore, $S$ is a del Pezzo $n$ surface $dP_n$ with $n \ge 2$
in which $h^{2,0}(S)=0$.
By the way, the Hirzebruch surfaces with degree larger than 2
satisfy $h^{2,0}(S)=0$ but do not define the fully 
consistent decoupled models~\cite{Beasley:2008dc, Beasley:2008kw}.

 To describe the spectrum, we have to study the gauge theory 
of the worldvolume on the seven-branes.  We 
start from the maximal supersymmetric gauge theory on
$\mathbb{R}^{3,1}\times \mathbb{C}^{2}$ and then replace
$\mathbb{C}^{2}$ with the K\"ahler surface $S$. In order to have
four-dimensional ${\cal N}=1$ supersymmetry, the
maximal supersymmetric gauge theory on $\mathbb{R}^{3,1}\times
\mathbb{C}^{2}$ should be twisted. It was shown that there exists an
unique twist preserving ${\cal N}=1$ supersymmetry in four
dimensions, and chiral matters can arise from the bulk $S$ or the
codimension-one curve $\Sigma$ in $S$ which is the intersection
between the observable seven-branes and 
the other seven-brane(s)~\cite{Beasley:2008dc, Beasley:2008kw}.

In order to have the matter fields on $S$,
we consider a non-trivial vector bundle on $S$ with
a structure group $H_S$ which is a subgroup of $G_S$. Then the gauge
group $G_S$ is broken down to $\Gamma_S\times H_S$, and the adjoint
representation ${\rm ad}(G_S)$ of the $G_S$ is decomposed as 
\begin{equation}
{\rm ad}(G_S)\rightarrow
{\rm ad}(\Gamma_S)\bigoplus {\rm ad}(H_S)\bigoplus_j(\tau_j,T_j)~.~\,
\end{equation}
Employing the vanishing theorem of the del Pezzo surfaces,
we obtain the numbers of the generations and anti-generations 
by calculating the zero modes of the Dirac operator on $S$
\begin{eqnarray}
 n_{\tau_j}~=~ -\chi (S, \mathbf{T_j})~,~~~ 
n_{\tau_j^*}~=~ -\chi (S, \mathbf{T_j}^*)~,~\,
\end{eqnarray}
where $\mathbf{T_j}$ is the vector bundle on $S$ whose 
sections transform in the representation $T_j$ of $H_S$,
and $\mathbf{T_j}^*$ is the dual bundle of $\mathbf{T_j}$.
In particular, when the $H_S$ bundle is a line bundle $L$,
we have
\begin{eqnarray}
n_{\tau_j}~=~-\chi (S, L^j)~=~
-\Big[1+\frac{1}{2}\big(\int_{S}c_{1}({L}^{j})c_{1}(S)+
\int_{S}c_{1}({L}^{j})^2\big)\Big]~.~\,
\label{EulerChar}
\end{eqnarray}
In order to preserve supersymmetry, the line bundle $L$ should satisfy
 the BPS equation~\cite{Beasley:2008dc}
\begin{equation}
J_{S}\wedge c_{1}(L)=0,\label{BPS}
\end{equation}
where $J_{S}$ is the K\"ahler form on $S$. Moreover,
the admissible supersymmetric line bundles on del Pezzo surfaces must
satisfy $c_{1}(L)c_{1}(S)=0$, thus,
 $n_{\tau_j}=n_{\tau_j^*}$ and only the vector-like particles
can be obtained. In short, we can not have the chiral matter fields
on the worldvolume of the observable seven-branes.

Interestingly, the chiral superfields can come from the intersections
between the observable seven-branes and the other 
seven-brane(s)~\cite{Beasley:2008dc, Beasley:2008kw}. 
Let us consider a stack of seven-branes with gauge group
$G_{S'}$ that wrap a codimension-one surface $S'$ in $B_3$.
The intersection of $S$ and $S'$ is a codimenion-one curve
(Riemann surface) $\Sigma$ in $S$ and $S'$, 
and the gauge symmetry on $\Sigma$
will be enhanced to $G_{\Sigma}$ 
where $G_{\Sigma}\supset G_{S}\times G_{S'}$.
On this curve, there exist chiral matters from 
the decomposition of the adjoint representation
${\rm ad}G_{\Sigma}$ of $G_{\Sigma}$ as follows
\begin{equation}
{\rm ad}G_{\Sigma}={\rm ad}G_{S}\oplus {\rm ad}G_{S'}\oplus_{k}
({ U}_{k}\otimes { U'}_{k})~.~\,
\end{equation}
Turning on the non-trivial gauge bundles on $S$ and $S'$ respectively
with structure groups $H_S$ and $H_{S'}$, we break the gauge 
group $G_S\times G_{S'}$ down to the commutant subgroup 
$\Gamma_{S}\times\Gamma_{S'}$. Defining 
$\Gamma \equiv \Gamma_{S}\times\Gamma_{S'}$ and 
$H \equiv H_{S}\times H_{S'}$,
we can decompose ${ U}\otimes { U'}$ into the irreducible
representations as follows
\begin{equation}
{ U}\otimes { U'}={\bigoplus}_{k}(r_{k}, {V}_{k}),
\end{equation}
where $r_{k}$ and ${ V}_{k}$ are the representations of $\Gamma$
and $H$, respectively. The light chiral fermions in the
representation $r_{k}$ are determined by the zero modes of the
Dirac operator on $\Sigma$. The net number of chiral superfields
 is given by
\begin{eqnarray}
N_{r_{k}}-N_{r^{*}_{k}}=\chi(\Sigma,K^{1/2}_{\Sigma}\otimes
{\mathbf{V}_{k}}),
\end{eqnarray}
where $K_{\Sigma}$ is the  restriction of
canonical bundle on the curve $\Sigma$, and
$\mathbf{V}_{k}$ is the vector bundle whose sections 
transform in the representation ${ V}_{k}$ of 
the structure group $H$. 

In the F-theory model building, we are interested in the 
models where $G_{S'}$ is $U(1)'$, and
$H_S$ and $H_{S'}$ are respectively $U(1)$
and $U(1)'$. Then the vector bundles on $S$ and $S'$ 
are line bundles $L$ and $L'$. The adjoint representation
${\rm ad}G_{\Sigma}$ of $G_{\Sigma}$ is 
decomposed into a direct sum
of the irreducible representations under the group
$\Gamma_S \times U(1) \times U(1)'$ that can be
denoted as $\mathbf{(r_j, q_j, q'_j)}$
\begin{equation}
{\rm ad}G_{\Sigma}={\rm ad}(\Gamma_S)
\oplus {\rm ad}G_{S'}\oplus_{j}
\mathbf{(r_j, q_j, q_j')}~.~\,
\end{equation}
The numbers of chiral supefields  in the representation 
$\mathbf{(r_j, q_j, q'_j)}$ and their Hermitian conjugates
on the curve $\Sigma$ are given by 
\begin{eqnarray}
N_{\mathbf{(r_j, q_j, q'_j)}} ~=~ h^0 (\Sigma, \mathbf{V}_j) ~,~~~
N_{\mathbf{({\bar r}_j, -q_j, -q'_j)}} ~=~ h^1(\Sigma, \mathbf{V}_j)~,~\,
\end{eqnarray}
where 
\begin{eqnarray}
\mathbf{V}_j~=~ K^{1/2}_{\Sigma} \otimes
{L}_{\Sigma}^{q_{j}}\otimes {L'}_{\Sigma}^{q'_{j}} ~,~\,
\end{eqnarray}
where $K^{1/2}_{\Sigma}$, ${L}_{\Sigma}^{r_{j}}$ and
${L'}_{\Sigma}^{q'_{j}}$ are the restrictions of
canonical bundle $K_S$, line bundles $L$ and $L'$ on the curve
$\Sigma$, respectively. In particular, if the
volume of $S'$ is infinite, $G_{S'}=U(1)'$ is decoupled.
And then the index $\mathbf{q'_j}$ can be ignored.

Using Riemann-Roch theorem, we obtain the net number of 
chiral supefields in the representation $\mathbf{(r_j, q_j, q'_j)}$
\begin{eqnarray}
N_{\mathbf{(r_j, q_j, q'_j)}}-
N_{\mathbf{({\bar r}_j, -q_j, -q'_j)}}~=~ 1-g+{\rm deg}(\mathbf{V}_j) ~,~\,
\end{eqnarray}
where $g$ is the genus of the curve $\Sigma$.

Moreover, we can obtain the Yukawa couplings 
at the triple intersections of
three curves $\Sigma_i$, $\Sigma_j$ and $\Sigma_k$ where
 the gauge group or the singularity type is enhanced further.
To have the triple intersections, the corresponding
homology classes  $[\Sigma_i]$, $[\Sigma_j]$ and $[\Sigma_k]$
of the curves $\Sigma_i$, $\Sigma_j$ and $\Sigma_k$ must satisfy
the following conditions
\begin{eqnarray}
[\Sigma_i] \cdot [\Sigma_j] > 0 ~,~~~
[\Sigma_i] \cdot [\Sigma_k] > 0 ~,~~~
[\Sigma_j] \cdot [\Sigma_k] > 0 ~.~\,
\label{FTYK-Con}
\end{eqnarray}

In this paper, we will construct the $SU(5)$ and $SO(10)$ models.
For simplicity, we will
consider $S$ to be the del Pezzo 8 surface $dP_8$ which is wrapped
by the observable seven-branes.


\section{$SU(5)$ Model}

In this Section, we will construct the $SU(5)$ model from F-theory.
First, let us briefly review the $SU(5)$ model and explain 
the convention. For convenience, 
we define the $U(1)_{Y}$ hypercharge generator in $SU(5)$ as follows
\bea 
T_{\rm U(1)_{Y}}={\rm diag} \left(-2, -2, -2,
 3,  3 \right)~.~\,
\label{u1y}
\eea
Under $SU(3)_C\times SU(2)_L \times U(1)_Y$ gauge symmetry, 
the $SU(5)$ representations
are decomposed as follows
\begin{eqnarray}
\mathbf{5} &=& \mathbf{(3, 1, -2)} \oplus \mathbf{(1, 2, 3)}~,~ \\
\mathbf{\overline{5}}  &=& 
\mathbf{(\overline{3}, 1, 2)} \oplus \mathbf{(1, 2, -3)}~,~ \\
\mathbf{10} &=& \mathbf{(3, 2, 1)} \oplus \mathbf{({\overline{3}}, 1, -4)}
\oplus \mathbf{(1, 1, 6)}~,~ \\
\mathbf{\overline{10}} &=& \mathbf{(\overline{3}, 2, -1)} 
\oplus \mathbf{(3, 1, 4)}
\oplus \mathbf{(1, 1, -6)}~,~ \\
\mathbf{24} &=&  \mathbf{(8, 1, 0)} \oplus  \mathbf{(1, 3, 0)} 
\oplus  \mathbf{(1, 1, 0)}  \oplus \mathbf{(3, 2, -5)} \oplus
 \mathbf{(\overline{3}, 2, 5)}~.~\,
\end{eqnarray}
There are three families of the SM fermions 
whose quantum numbers under $SU(5)$ are
\bea
F_i=\mathbf{10},~ {\overline f}_i={\mathbf{\bar 5}},~
N^c_i={\mathbf{1}}~,~
\label{SU(5)-smfermions}
\eea
where $i=1, 2, 3$, and $N^c_i$ are the right-handed neutrinos. 
The SM particle assignments in $F_i$ and ${\bar f}_i$  are
\bea
F_i=(Q_i, U^c_i, E^c_i)~,~{\overline f}_i=(D^c_i, L_i)~,~
\label{SU(5)-smparticles}
\eea
where $Q_i$ and $L_i$ are respectively  the left-handed
quark and lepton doublets, and $U^c_i$, $D^c_i$, and $E^c_i$ are 
 the right-handed up-type quarks,
down-type quarks, and charged leptons, respectively.
In our model building, we will introduce the SM singlet
fields $\overline{E}^c_i$ from $\mathbf{\overline{10}}$ whose SM quantum 
numbers are $\mathbf{(1, 1, -6)}$. Especially, $\overline{E}^c_i$ are
Hermitian conjugate of $E_i^c$, and then they can form
vector-like particles.

To break the electroweak gauge symmetry, we introduce one pair
of Higgs fields whose quantum numbers under $SU(5)$ are
\bea
h={\mathbf{5}}~,~~~{\overline h}={\mathbf{\bar {5}}}~.~\,
\label{SU(5)-1-Higgse}
\eea
Explicitly, we denote the Higgs particles as follows
\bea
h=(D_h, D_h, D_h, H_u)~,~
{\overline h}=({\overline {D}}_{\overline h}, {\overline {D}}_{\overline h},
{\overline {D}}_{\overline h}, H_d)~,~\,
\label{SU(5)-3-Higgse3}
\eea
where $H_u$ and $H_d$ are one pair of the Higgs doublets in the 
Minimal Supersymmetric Standard Model (MSSM).
In $SU(5)$ model, the superpotential for 
the SM fermion Yukawa couplings and the mass terms of
the Higgs fields and right-handed neutrinos is 
\bea 
{ W}_{\rm Yukawa} = y_{ij}^{U}
F_i F_j h + y_{ij}^{D E} F_i  {\overline f}_j {\overline h}
+ y_{ij}^{\nu} N^c_i  {\overline f}_j h + \mu h {\overline h}
+m^N_{ij} N_i^c N_j^c~,~\,
\label{SU(5)-SP}
\eea
where $y_{ij}^{U}$, $y_{ij}^{D E}$, $y_{ij}^{\nu}$ 
are Yukawa couplings, 
$\mu$  is the bilinear Higgs mass term, and $m^N_{ij}$
are the Majorana masses for the right-handed neutrinos.
So, we have the Yukawa coupling unification for
the bottom quark and tau lepton.

Because we will construct $SU(5)$ model,  we choose 
$G_S=SU(5)$ and $H_S=U(1)_Y$. The SM fermions 
$\overline{f}_i$, and  the 
Higgs fields $h$ and $\overline{h}$ are on the curves 
where the $SU(5)$ gauge symmetry is enhanced to $SU(6)$. 
Under $SU(5)\times U(1)$, the adjoint representation of $SU(6)$
is decomposed as follows
\begin{eqnarray}
\mathbf{35} &=& \mathbf{(24, 0)} \oplus \mathbf{(1, 0)}
\oplus \mathbf{(5, 6)} \oplus \mathbf{({\overline{5}}, -6)}~.~\,
\end{eqnarray}
Also, the SM fermions $F_i$ are on the curves where
the $SU(5)$ gauge symmetry is enhanced to $SO(10)$. 
Under $SU(5)\times U(1)$, the adjoint representation of $SO(10)$
is decomposed as follows 
\begin{eqnarray}
\mathbf{45} &=& \mathbf{(24, 0)} \oplus \mathbf{(1, 0)}
\oplus \mathbf{(10, 4)} \oplus \mathbf{({\overline{10}}, -4)}~.~\,
\end{eqnarray}

In addition, the SM fermion Yukawa couplings 
$ y_{ij}^{D E} F_i  {\overline f}_j {\overline h}$ arise from
the triple intersection where the gauge symmetry is enhanced to 
$SO(12)$. Under $SU(5)\times U(1)_1\times U(1)_2$,  the 
adjoint representation of $SO(12)$ is decomposed as follows 
\begin{eqnarray}
\mathbf{66} &=& \mathbf{(24, 0, 0)} \oplus \mathbf{(1, 0, 0)}
\oplus \mathbf{(1, 0, 0)} \oplus  \mathbf{(5, 2, 2)}
  \oplus  \mathbf{(5, 2, -2)} 
\oplus \mathbf{({\overline{5}}, -2, 2)}
\nnb \\&&
\oplus \mathbf{({\overline{5}}, -2, -2)}
\oplus \mathbf{(10, 4, 0)} \oplus \mathbf{({\overline{10}}, -4,0)}~.~\,
\end{eqnarray}
We denote the $U(1)_1\times U(1)_2$ generators in $SO(12)$
as $t_1$ and $t_2$. We emphasize that the field theory description
in terms of $SU(5)\times U(1)_1\times U(1)_2$ is valid only within
a small open neighborhood of the $SO(12)$ point.
Thus, to preserve the $U(1)_1\times U(1)_2$ gauge or global 
symmetry for the down-type quark and lepton Yukawa couplings
in the small open neighborhood of the $SO(12)$ point, 
we obtain that  
the SM fermions $F_i$, ${\overline f}_j$, and Higgs fields
${\overline h}$ should localize
on the curves $t_1=0$, $t_1+t_2=0$ (or $t_1-t_2=0$), 
and $t_1-t_2=0$ (or $t_1+t_2=0$), respectively.

Also, the SM fermion Yukawa couplings 
$ y_{ij}^{U} F_i F_j h $ arise from
the triple intersection where the gauge symmetry is enhanced to 
$E_6$. Because in our models $Q_i$ and $U_i^c$ are not in
the same  $F_i$ multiplets, we denote these Yukawa couplings as 
$y_{ij}^{U} \mathbf{10}_i \mathbf{10}'_j h$.
Under $SU(5)\times U(1)_a\times U(1)_b$,  the 
adjoint representation of $E_6$ is decomposed as follows 
\begin{eqnarray}
\mathbf{78} &=& \mathbf{(24, 0, 0)} \oplus \mathbf{(1, 0, 0)}
\oplus \mathbf{(1, 0, 0)} \oplus \mathbf{(1, 5, 3)} 
 \oplus \mathbf{(1, -5, -3)} 
\oplus  \mathbf{(5, -3, 3)} 
\nnb \\&&
\oplus \mathbf{({\overline{5}}, 3, -3)}
\oplus \mathbf{(10, -1, -3)} \oplus \mathbf{({\overline{10}}, 1,3)}
\oplus \mathbf{(10, 4, 0)} \oplus \mathbf{({\overline{10}}, -4,0)}~.~\,
\end{eqnarray}
We denote the $U(1)_a\times U(1)_b$ generators in $E_6$
as $t_a$ and $t_b$. We emphasize that the field theory description
in terms of $SU(5)\times U(1)_a\times U(1)_b$ is valid only within
a small open neighborhood of the $E_6$ point.
Thus, to preserve the $U(1)_a\times U(1)_b$ gauge or global 
symmetry for the SM up-type quark Yukawa couplings
in the small open neighborhood of the $E_6$ point, we obtain that  
the SM fermions $\mathbf{10}_i$, $\mathbf{10}'_j$, 
and Higgs field
${h}$ should localize
on the curves $t_a=0$ (or $t_a+3t_b=0$), 
$t_a+3t_b=0$ (or $t_a=0$),  
and $t_a-t_b=0$, respectively.
Therefore, to have the up-type quark Yukawa couplings
which are invariant under $U(1)_a\times U(1)_b$, 
we obtain that $Q_i$ and $U_j^c$ should localize on 
the matter curves $t_a=0$ (or $t_a+3t_b=0$), and  
$t_a+3t_b=0$ (or $t_a=0$), respectively.

By the way, the simplest possibility is to put one complete 
 $\mathbf{10}$ of the SM fermions on the 
matter curve $t_a=0$ (or $t_a+3t_b=0$) 
and the remaining two complete $\mathbf{10}$s of the SM fermions
on the matter curve $t_a+3t_b=0$ (or $t_a=0$), respectively. However,
because the first  $\mathbf{10}$  can only couple to the second and third 
 $\mathbf{10}$s but not to
itself, and similarly, the second and third $\mathbf{10}$s 
can couple only to the first  $\mathbf{10}$, we obtain
the up-type quark Yukawa matrix 
\begin{equation}
y^U \sim \begin{pmatrix}0 & \ast & \ast \\ \ast & 0 & 0
\\ \ast & 0 & 0 \end{pmatrix}~.~
\end{equation}
Note that there are  two massless up-type quarks,
this possibility is disfavored from the phenomenological point of 
view. Thus, we put the left-handed quark doublets
 $Q_i$ and the right-handed up-type quarks  $U_j^c$ on
the different matter curves.


\begin{table}[htb]
\begin{center}
\begin{tabular}{|c|c|c|c|c|c|}
\hline
${\rm Particles }$ & ${\rm Curve}$ & ${\rm Class}$ & $g_{\Sigma}$ &
$L_{\Sigma}$ & $L_{\Sigma}^{\prime n}$\\\hline
$H_u$ & $\Sigma_{h}$ & $2H-E_{1}-E_{3}$ & $0$ &
$\mathcal{O}_{\Sigma_{h}}(1)^{1/5}$ & $\mathcal{O}_{\Sigma_{h}}
(1)^{2/5}$\\\hline
$H_d$ & $\Sigma_{\overline{h}}$ & $2H-E_{2}-E_{3}$ & $0$ &
$\mathcal{O}_{\Sigma_{\overline{h}}}(-1)^{1/5}$ & 
$\mathcal{O}_{\Sigma_{\overline{h}}}(-1)^{2/5}$\\\hline
$\overline{f}_i$ & $\Sigma_{f}$ & $2H$ & $0$ & 
$\mathcal{O}_{\Sigma_{f}}$ & $\mathcal{O}_{\Sigma_{f}}(-3p^{\prime})$\\\hline
$\left(E_{2i-1}^c,~E_{2i}^c,~Q_i\right)$ &
$\Sigma_{Qi}$ & $2H-E_1-E_{i+3}$ & 0 &
$\mathcal{O}_{\Sigma_{Qi}}(1)^{1/5}$ &   $\mathcal{O}_{\Sigma_{Qi}}
(1)^{4/5}$\\\hline
$\left(U_i^c,~\overline{E}^c_i \right)$ &
$\Sigma_{Ui}$ & $2H-E_2-E_{i+3}$ & 0 &
$\mathcal{O}_{\Sigma_{Ui}}(-1)^{1/5}$ &   $\mathcal{O}_{\Sigma_{Ui}}
(1)^{1/5}$\\\hline
\end{tabular}
\end{center}
\caption{ The particle curves
and the gauge bundle assignments for each curve in the $SU(5)$ model.
Here, $i$ is the SM fermion family index, {\it i.e.}, $i=1,~2,~3$.}
\label{tab:SU5}
\end{table}


In our $SU(5)$ model, we choose the line bundle 
$L=\mathcal{O}_{S}(E_{1}-E_{2})^{1/5}$, and
 present the particle curves with homology classes and the 
gauge bundle assignments for each curve  in Table~\ref{tab:SU5}.
We do not have the vector-like particles from the bulk of
the observable seven-branes since $\chi (S, L^5)=0$.
Note that the Higgs triplets in $h$ and $\overline{h}$ do not
have zero modes, we solve the doublet-triplet splitting problem.
Also, we have six $E_i^c$ fields and three $\overline{E}_i^c$ fields.
Because they are vector-like, we indeed have three chiral $E_i^c$ fields.
In addition, we assume that  the SM fermion and Higgs
 curves  $\Sigma_{Qi}$,
 $\Sigma_{Ui}$, and $\Sigma_h$, intersect at 
one point in $S$ where the gauge symmetry is enhanced to $E_6$. 
And we introduce the SM singlet Higgs fields
$X_k$ and $\overline{X}_k$ respectively
with $SU(5)\times U(1)_a\times U(1)_b$ quantum numbers
$\mathbf{(1, 5, 3)}$ and $\mathbf{(1, -5, -3)}$ at this $E_6$ point
from the intersections of the other seven-branes.
Moreover, we assume that  the SM fermion and Higgs curves  
$\Sigma_{Qi}$, $\Sigma_{f}$, and $\Sigma_{\overline{h}}$,
 intersect at the other
  point in $S$ where the gauge symmetry is enhanced to $SO(12)$.
In order to have the SM fermion Yukawa couplings which are invariant
under $U(1)_1\times U(1)_2$ and $U(1)_a\times U(1)_b$ symmetries,
we choose that the Higgs curves $\Sigma_h$ and $\Sigma_{\overline{h}}$
satisfy $t_1-t_2=0$ in the small open neighborhood of the $SO(12)$ point
 and $t_a-t_b=0$ in the small open neighborhood of the $E_6$ point,
the SM fermion curve $\Sigma_f$ satisfies $t_1+t_2=0$ 
in the small open neighborhood of the $SO(12)$ point, 
the SM fermion curve $\Sigma_{Qi}$ satisfies
 $t_1=0$ in the small open neighborhood of the $SO(12)$ point
 and $t_a=0$ in the small open neighborhood of the $E_6$ point,
 and the SM fermion curve $\Sigma_{Ui}$ satisfies $t_a+3t_b=0$  
in the small open neighborhood of the $E_6$ point.
Therefore, the superpotential in our $SU(5)$ model is
\bea 
{ W_{SSM}}&=&
y_{ij}^{D} D^c_i Q_j H_d+ y_{ij}^{U } U^c_i Q_j H_u
+ y_{ij}^{E} E^c_i L_j H_d+ \lambda_1^{ijk} \overline{X}_i E^c_j \overline{E}_k^c
 +  \mu H_d H_u~,~\, 
\label{SU(5)-poten}
\eea
where $y_{ij}^{D}$, $y_{ij}^{U }$, $y_{ij}^{E}$, and $\lambda_1^{ijk}$
are Yukawa couplings. $\mu$ term can be generated via the instanton
effects or the Higgs mechanism since $H_u$ and $H_d$
can couple to the SM singlet Higgs fields from the intersection of the other
seven-branes. Interestingly, we can also have the Yukawa coupling unification 
for the bottom quark and tau lepton 
at the GUT scale. Although the SM fermion Yukawa matrices
are rank one, the SM fermion
masses and mixings can be generated in the presence
of $H$-fluxes~\cite{Cecotti:2009zf}.
Also, after $\overline{X}_i$ obtain the vacuum
expectation values (VEVs),  three linear combinations of the
six $E_i^c$ fields and three $\overline{E}_i^c$ fields will obtain
the vector-like masses and then decouple at the high scale. Thus, only three linear
combinations of six $E_i^c$ fields will be massless,
and then we have the supersymmetric Standard Model at the low energy.

In the $SU(5)$ models, the tree-level gauge kinetic functions with the $U(1)_Y$
flux contributions at the string scale
 are~\cite{Donagi:2008kj, Blumenhagen:2008aw}
\begin{eqnarray}
f_{SU(3)_C}= \tau_o -  
{1\over 2} \tau \int_S  c_1^2(L_o) ~,~\,
\end{eqnarray}
\begin{eqnarray}
f_{SU(2)_L}= \tau_o -  
{1\over 2} \tau \int_S \left( c_1^2(L_o) + c_1^2(L^5) \right)~,~\,
\end{eqnarray}
\begin{eqnarray}
f_{U(1)_Y}= \tau_o -  
{1\over 2} \tau \int_S \left( c_1^2(L_o) + {3\over 5} c_1^2(L^5) \right)~,~\,
\end{eqnarray}
where $\tau_o$ is the original gauge kinetic function of $SU(5)$,
and $L_o$ is the restriction of the line bundle on the
internal Calabi-Yau manifold.
Thus, we obtain the SM gauge coupling relation at the string 
scale~\cite{Donagi:2008kj, Blumenhagen:2008aw}
\begin{eqnarray}
\alpha_1^{-1}-\alpha_3^{-1} &=& {3\over 5} (\alpha_2^{-1}-\alpha_3^{-1}) ~,~
\alpha_3^{-1} < \alpha_1^{-1} < \alpha_2^{-1} ~,~ \,
\end{eqnarray}
where $\alpha_1$, $\alpha_2$, and $\alpha_3$ are the gauge couplings for
the $U(1)_Y$, $SU(2)_L$, and $SU(3)_C$, respectively.
Interestingly, in our $SU(5)$ model, we have three linear combinations of the
six $E_i^c$ fields and three $\overline{E}_i^c$ fields that will obtain
the heavy vector-like masses. If we consider these vector-like 
particles as heavy threshold corrections to the RGE running  
of the SM gauge couplings, we can explain the
 SM gauge couplings at the string scale.
The detailed discussions will be given elsewhere.



\section{$SO(10)$ Models}

In this Section, we will construct two $SO(10)$ models from F-theory.
Turning on the $U(1)_{B-L}$ flux, we can break the $SO(10)$ gauge
symmetry down to the $SU(3)_C \times SU(2)_L \times SU(2)_R \times U(1)_{B-L}$
gauge symmetry.  For convenience, 
we define the $U(1)_{B-L}$ generator in $SU(4)_C$  of  $SO(10)$ as follows 
\bea 
T_{\rm U(1)_{B-L}}={\rm diag} \left(1, 1, 1,
 -3 \right)~.~\,
\label{u1y}
\eea
Under $SU(3)_C\times SU(2)_L \times SU(2)_R \times U(1)_{B-L}$ gauge symmetry, 
the $SO(10)$ representations are decomposed as follows
\begin{eqnarray}
\mathbf{10} &=& \mathbf{(1, 2, 2, 0)} \oplus 
\mathbf{(3, 1, 1, -2)} \oplus 
\mathbf{({\overline{3}}, 1, 1, 2)} ~,~ \\
\mathbf{{16}} &=&  \mathbf{(3, 2, 1, 1)} 
\oplus  \mathbf{(1, 2, 1, -3)} \oplus 
\mathbf{(\overline{3}, 1, 2, -1)} 
\oplus  \mathbf{(1, 1, 2, 3)}~,~\\
\mathbf{45} &=&  \mathbf{(8, 1, 1, 0)} \oplus  \mathbf{(1, 3, 1, 0)} 
\oplus  \mathbf{(1, 1, 3, 0)}  \oplus  \mathbf{(1, 1, 1, 0)} 
\oplus \mathbf{(3, 1, 1, 4)} 
\nnb \\&&
\oplus  \mathbf{({\overline{3}}, 1, 1, -4)}
\oplus \mathbf{(3, 2, 2, -2)} 
\oplus  \mathbf{({\overline{3}}, 2, 2, 2)}
~.~\,
\end{eqnarray}

Under the $SU(3)_C\times SU(2)_L\times SU(2)_R\times U(1)_{B-L}$
gauge symmetry, the SM fermions are
\bea
Q_i~=~ \mathbf{(3, 2, 1, 1)}~,~~ Q^R_i~=~\mathbf{(\overline{3}, 1, 2, -1)} ~,~~
L_i~=~  \mathbf{(1, 2, 1, -3)} ~,~~ L^R_i~=~\mathbf{(1, 1, 2, 3)}~,~\,
\label{SO(10)-smfermions}
\eea
where $i=1, 2, 3$.
The SM particle assignments in $Q^R_i$, and $L^R_i$  are
\bea
Q^R_i~=~(U^c_i, D^c_i)~,~~ L^R_i~=~(E_i^c, N_i^c)~.~\,
\label{SO(10)-smparticles}
\eea

To break the $SU(2)_R\times U(1)_{B-L}$ gauge symmetry down to
$U(1)_Y$ and to break 
the electroweak gauge symmetry, we introduce the following
 Higgs fields whose quantum numbers
under the $SU(3)_C\times SU(2)_L\times SU(2)_R\times U(1)_{B-L}$
gauge symmetry are
\bea
\Phi~=~\mathbf{(1, 1, 2, 3)}~,~~\overline{\Phi} = \mathbf{(1, 1, 2, -3)}
~,~~ H~=~ \mathbf{(1, 2, 2, 0)}~,~\,
\label{SO(10)-1-Higgse}
\eea
where $\Phi$ and $\overline{\Phi}$ are the Higgs fields to break
the $SU(2)_R\times U(1)_{B-L}$ gauge symmetry, and
$H$ contains both $H_u$ and $H_d$ in the MSSM.

In the $SU(3)_C\times SU(2)_L\times SU(2)_R\times U(1)_{B-L}$ model, 
the superpotential for  the SM fermion Yukawa couplings and Higgs mass
term is 
\bea 
{ W}_{\rm Yukawa} = y_{ij}^{UD} Q_i H Q^R_j + y_{ij}^{E\nu} L_i H L^R_j
+\mu H H~,~\,
\label{SO(10)-SP}
\eea
where $y_{ij}^{UD}$ and $y_{ij}^{E\nu}$
are Yukawa couplings,  and $\mu$  is the bilinear Higgs mass term.
Thus, we obtain  the Yukawa coupling unification for 
the SM quarks and  the Yukawa coupling unification for 
the SM leptons and neutrinos at the GUT scale.

 In our model building, we will also introduce the additional vector-like
particles $U'_i$ and $U^{\prime c}_i$ from the bulk $S$,  and the particles 
$\overline{Q}_i$, $\overline{Q}^{R}_i$, $\overline{L}_i$
and  $\overline{L}^{R}_i$ from $\mathbf{\overline{16}}$, whose
quantum numbers under the $SU(3)_C\times SU(2)_L\times SU(2)_R\times U(1)_{B-L}$
gauge symmetry are
\bea
U'_i&=&\mathbf{(3, 1, 1, 4)}~,~~ U^{\prime c}_i~=~ \mathbf{({\overline{3}}, 1, 1, -4)}~,~~ 
\overline{Q}_i~=~\mathbf{(\overline{3}, 2, 1, -1)}~,~~
\nnb \\
\overline{Q}^{R}_i &=& \mathbf{({3}, 1, 2, 1)}~,~~
\overline{L}_i ~=~\mathbf{(1, 2, 1, 3)}~,~~
\overline{L}^{R}_i~=~\mathbf{(1, 1, 2,  -3)}~.~\,
\eea

Because we will construct the $SO(10)$ models, 
we choose $G_S=SO(10)$ and 
$H_S=U(1)_{B-L}$.  The bidoublet Higgs field $H$ is on the curve 
where the $SO(10)$ gauge symmetry is enhanced to $SO(12)$. 
Under $SO(10)\times U(1)$, the adjoint representation of $SO(12)$
is decomposed as follows
\begin{eqnarray}
\mathbf{66} &=& \mathbf{(45, 0)} \oplus \mathbf{(1, 0)}
\oplus \mathbf{(10, 2)} \oplus \mathbf{({{10}}, -2)}~.~\,
\end{eqnarray}
All the other fields in our models are on the curves where
the $SO(10)$ gauge symmetry is enhanced to $E_6$. 
Under $SO(10)\times U(1)$, the adjoint representation of $E_6$
is decomposed as follows 
\begin{eqnarray}
\mathbf{78} &=& \mathbf{(45, 0)} \oplus \mathbf{(1, 0)}
\oplus \mathbf{(16, -3)} \oplus \mathbf{({\overline{16}}, 3)}~.~\,
\end{eqnarray}

In addition, the SM fermion Yukawa couplings in $SO(10)$ models arise
from the triple intersections where the gauge symmetry is enhanced
to $E_7$.
Under $SO(10)\times U(1)_a\times U(1)_b$,  the 
adjoint representation of $E_7$ is decomposed as follows 
\begin{eqnarray}
\mathbf{133} &=& \mathbf{(45, 0, 0)} \oplus \mathbf{(1, 0, 0)}
\oplus \mathbf{(1, 0, 0)} \oplus \mathbf{(1, 0, 2)} 
 \oplus \mathbf{(1, 0, -2)} 
\oplus  \mathbf{(10, 2, 0)} \nnb \\&&
\oplus  \mathbf{(10, -2, 0)} 
\oplus  \mathbf{(16, -1, 1)}
\oplus  \mathbf{(16, -1, -1)}
 \oplus \mathbf{({\overline{16}}, 1, 1)}
 \oplus \mathbf{({\overline{16}}, 1, -1)}~.~\,
\end{eqnarray}
We denote the $U(1)_a\times U(1)_b$ generators in $E_7$
as $t_a$ and $t_b$.
We emphasize that the field theory description
in terms of $SO(10)\times U(1)_a\times U(1)_b$ is valid only within
a small open neighborhood of the $E_7$ point.
Thus, to preserve the $U(1)_a\times U(1)_b$ gauge or global 
symmetry for the SM fermion Yukawa couplings
in the small open neighborhood of the $E_7$ point, we obtain that  
the bidoublet Higgs field $H$ localize on the curve
$t_a=0$, the SM quarks $Q_i$ and $Q^R_i$ should localize 
respectively on 
the matter curves $t_a+t_b=0$ (or $t_a-t_b=0$) and  
$t_a-t_b=0$ (or $t_a+t_b=0$), and the SM leptons $L_i$
and $L^R_i$ should localize 
respectively on 
the matter curves $t_a-t_b=0$ (or $t_a+t_b=0$) and  
$t_a+t_b=0$ (or $t_a-t_b=0$).

Moreover, the Higgs Yukawa couplings in $SO(10)$ models arise
from the Higgs curve intersection where the gauge symmetry 
is enhanced to $SO(14)$. Under $SO(10)\times U(1)_1\times U(1)_2$,  
the adjoint representation of $SO(14)$ is decomposed as follows 
\begin{eqnarray}
\mathbf{91} &=& \mathbf{(45, 0, 0)} \oplus \mathbf{(1, 0, 0)}
\oplus \mathbf{(1, 0, 0)} \oplus \mathbf{(10, 2, 0)} 
 \oplus \mathbf{(10, -2, 0)} 
\oplus  \mathbf{(10, 0, 2)} \nnb \\&&
\oplus \mathbf{(1, 2, 2)} \oplus \mathbf{(1, -2, 2)}
\oplus  \mathbf{(10, 0, -2)}
\oplus \mathbf{(1, 2, -2)} \oplus \mathbf{(1, -2, -2)}~.~\,
\end{eqnarray}
We denote the $U(1)_1\times U(1)_2$ generators in $SO(14)$
as $t_1$ and $t_2$. We emphasize that the field theory description
in terms of $SO(10)\times U(1)_1\times U(1)_2$ is valid only within
a small open neighborhood of the $SO(14)$ point.
And we assume that
 the bidoublet Higgs field $H$ localize 
on the curve $t_1=0$.

In this Section, we will take the line bundle 
as $L=\mathcal{O}_{S}(E_{1}-E_{2})^{1/2}$.
Note that $\chi(S, L^4)=3$, we  have three pairs of
 vector-like particles $U'_i$ and $U^{\prime c}_i$ from 
the bulk of the observable seven-branes.
These vector-like particles $U'_i$ and $U^{\prime c}_i$
 can obtain
masses via the instanton effects. Also, they can couple to 
the SM singlet Higgs fields
from the intersections of the other seven-branes, and then obtain
masses from the Higgs mechanism. For simplicity, in this paper, we will
assume that the vector-like particles $U'_i$ and $U^{\prime c}_i$
 have masses around the GUT scale and then decouple. 
In the following two subsections,
we will present two $SO(10)$ models.

Before we construct the $SO(10)$ models, let us consider the $U(1)_{B-L}$
flux contributions to the SM gauge couplings at the string scale. 
For $G=SO(10)$ gauge group, 
the generators $T^a$ of $SO(10)$ are imaginary
antisymmetric $10 \times 10$ matrices.  In terms of the $2\times 2$
identity matrix $\sigma_0$ and the Pauli matrices $\sigma_i$, they can
be written as tensor products of $2\times 2$ and $5 \times 5$
matrices, $(\sigma_0, \sigma_1, \sigma_3) \otimes A_5$ and $\sigma_2
\otimes S_5$ as a complete set, where $A_5$ and $S_5$ are the $5\times
5$ real anti-symmetric and symmetric matrices~\cite{Huang:2004ui}.  
The generators for $SU(4)_C\times SU(2)_L\times SU(2)_R$
  are~\cite{Huang:2004ui}
\begin{eqnarray}
&& (\sigma_0, \sigma_1, \sigma_3)  \otimes A_3\,, \quad 
(\sigma_0, \sigma_1, \sigma_3) \otimes A_2\,,  \nonumber \\
&& \sigma_2 \otimes S_3\,, \quad \sigma_2 \otimes S_2\, ,~\,
\end{eqnarray}
where $A_3$ and $S_3$ are respectively the diagonal blocks of $A_5$
and $S_5$ that have indices 1, 2, and 3, while the diagonal blocks
$A_2$ and $S_2$ have indices 4 and 5. In addition, 
the generator for the $U(1)_{B-L}$ is
\begin{eqnarray}
T_{B-L} = {1\over {\sqrt 3}} \sigma_2\otimes {\rm diag}(1, 1, 1, 0, 0) ~.~\,
\end{eqnarray}

The flux contributions to the gauge couplings can be computed 
by dimensionally reducing the Chern-Simons action of the
observable seven-branes wrapping on $S$
\begin{eqnarray}
S_{\rm CS} &=& \mu_7 \int_{S\times \mathbb{R}^{3,1}} a \wedge {\rm tr}(F^4) ~.~\,
\end{eqnarray}
In our $SO(10)$ models, we choose  the $U(1)_{B-L}$ flux as follows
\begin{eqnarray}
\langle F_{\rm U(1)_{B-L}} \rangle &=& {1\over 2} V_{U(1)_{B-L}} \sigma_2\otimes 
{\rm diag}(1, 1, 1, 0, 0)  ~.~\,
\end{eqnarray}
Let us noramlize  the $SO(10)$ generators $T^a$ as 
${\rm Tr}(T^aT^b)=2\delta_{ab}$.
Then, we obtain the tree-level gauge kinetic functions with the $U(1)_{B-L}$
flux contributions for $SU(3)_C$, $U(1)_{B-L}$, $SU(2)_L$, and $SU(2)_R$
gauge symmetries at the string scale 
\begin{eqnarray}
f_{SU(3)_C} = f_{U(1)_{B-L}}=
 \tau_o -  
{1\over 2} \tau \int_S  c_1^2(L^2) \equiv \tau_o + \tau ~,~\,
\end{eqnarray}
\begin{eqnarray}
f_{SU(2)_L}= f_{SU(2)_R} = \tau_o ~,~\,
\end{eqnarray}
where $\tau_o$ is the original gauge kinetic function of $SO(10)$.

We will break the $SU(2)_R\times U(1)_{B-L}$ gauge symmetry
at the string scale by Higgs mechanism in our $SO(10)$
models. Thus, we obtain the gauge kinetic function
for $U(1)_Y$ 
\begin{eqnarray}
 f_{U(1)_{Y}}= {3\over 5} f_{SU(2)_R} + {2\over 5} f_{U(1)_{B-L}}=
\tau_o + {2\over 5} \tau ~.~\,
\end{eqnarray}
Therefore, we obtain
the SM gauge coupling relation at the string scale
\begin{eqnarray}
\alpha_1^{-1}-\alpha_3^{-1} &=& {3\over 5} (\alpha_2^{-1}-\alpha_3^{-1}) ~,~
\alpha_2^{-1} < \alpha_1^{-1} < \alpha_3^{-1} ~.~ \,
\end{eqnarray}
We emphasize that although the SM gauge coupling relation at the string scale
in the $SO(10)$  models is the same
as that in the $SU(5)$ model, the order of the SM gauge couplings 
($\alpha_i^{-1}$) from small to large is reversed.


\subsection{Type I $SO(10)$ Model}


\begin{table}[htb]
\begin{center}
\begin{tabular}{|c|c|c|c|c|c|}
\hline
${\rm Particles }$ & ${\rm Curve}$ & ${\rm Class}$ & $g_{\Sigma}$ &
$L_{\Sigma}$ & $L_{\Sigma}^{\prime n}$\\\hline
$\left(H,~H'\right)$ & $\Sigma_{H}\text{ (pinched)}$ 
& $3H-E_{1}-E_{2}$ & $1$ &
$\mathcal{O}_{\Sigma_{H}}(p_1-p_2)^{1/2}$ & $\mathcal{O}_{\Sigma_{H}}$\\\hline
$\left(\Phi,~\overline{\Phi}\right)$ 
& $\Sigma_{\Phi}\text{ (pinched)}$ & $3H-E_{1}-E_{2}-E_8$ & $1$ &
$\mathcal{O}_{\Sigma_{\Phi}}(p'_1-p'_2)^{1/2}$ & 
$\mathcal{O}_{\Sigma_{\Phi}}(p'_1-p'_2)^{3/2}$\\\hline
$\left(L^R_{2i-1},~L^R_{2i},~Q_i,~\overline{L}_i\right)$ &
$\Sigma_{Qi}$ & $2H-E_1-E_{i+2}$ & 0 &
$\mathcal{O}_{\Sigma_{Qi}}(1)^{1/2}$ &   $\mathcal{O}_{\Sigma_{Qi}}
(-1)^{1/2}$\\\hline
$\left( L_{2i-1},~L_{2i},~Q^R_i,~\overline{L}_i^{R}\right)$ &
$\Sigma_{QRi}$ & $2H-E_2-E_{i+2}$ & 0 &
$\mathcal{O}_{\Sigma_{QRi}}(-1)^{1/2}$ &   $\mathcal{O}_{\Sigma_{QRi}}
(-1)^{1/2}$\\\hline
\end{tabular}
\end{center}
\caption{ The particle curves
and the gauge bundle assignments for each curve in the Type I $SO(10)$ model.
Here,  $i=1,~2,~3$.}
\label{SO(10)-I}
\end{table}



In Type I $SO(10)$ model, we present the particle curves with 
homology classes and the gauge bundle assignments for each curve  
in Table~\ref{SO(10)-I}. Note that the Higgs triplets in $\mathbf{10}$
of $SO(10)$ do not have zero modes, we solve the doublet-triplet
splitting problem. Also, we have six $L_i$ fields,  six $L_i^R$ fields,
three $\overline{L}_i$ fields, and three $\overline{L}_i^{R}$ fields.
Because $(L_i,~\overline{L}_i)$ and $(L_i^R, ~\overline{L}_i^{R})$ 
 are vector-like, the net numbers of chiral $L_i$  and 
$L_i^R$ superfields are three.
And we have two bidoublet Higgs fields $H$ and $H'$, which respectively
come from the $\mathbf{(10, 2)}$ and $\mathbf{(10, -2)}$ of 
the decompositions of the $SO(12)$ adjoint representation.

In addition, we assume that  the SM fermion and Higgs curves 
$\Sigma_{Qi}$,  $\Sigma_{QRi}$ and $\Sigma_H$ intersect
at one point in $S$ where the gauge symmetry is enhanced to $E_7$.
And we introduce the SM singlet Higgs fields
$X_k$ and $\overline{X}_k$ respectively 
with $SO(10)\times U(1)_a\times U(1)_b$ 
quantum numbers $\mathbf{(1, 0, 2)}$ and $\mathbf{(1, 0, -2)}$
 at this $E_7$ point from the intersection of the other seven-branes.
Moreover, we  assume that   the SM fermion and Higgs curves 
$\Sigma_{Qi}$ and $\Sigma_{\Phi}$ intersect
at the other point in $S$ where the gauge symmetry is enhanced to $E_7$
as well. 
And we introduce the SM singlet fields
$X'_k$ and $\overline{X}'_k$ respectively 
with $SO(10)\times U(1)_a\times U(1)_b$ 
quantum numbers $\mathbf{(1, 0, 2)}$ and $\mathbf{(1, 0, -2)}$
 at this $E_7$ point from the intersection of the other seven-branes. 
In order to have the SM fermion Yukawa couplings which are invariant
under $U(1)_a\times U(1)_b$ symmetry,
we choose that in the small open neighborhoods of these $E_7$ points,
 the Higgs curve $\Sigma_H$ 
satisfies $t_a=0$, the Higgs curve  $\Sigma_{\Phi}$ 
satisfies $t_a-t_b=0$, 
the SM fermion curve $\Sigma_{Qi}$ satisfies $t_a+t_b=0$, and 
the SM fermion curve $\Sigma_{QRi}$ satisfies $t_a-t_b=0$.
Futhermore,
 the Higgs curve $\Sigma_H$ 
is pinched at a point, where the gauge symmetry is enhanced to $SO(14)$.
We assume that the Higgs curve $\Sigma_H$ 
satisfies $t_1=0$ in the small open neighborhood of the $SO(14)$ point.
And we can introduce
the SM singlet Higgs fields $\phi^{++}_i$, $\phi^{+-}_i$, $\phi^{-+}_i$,
and $\phi^{--}_i$ respectively with $SO(10)\times U(1)_1\times U(1)_2$ 
quantum numbers $\mathbf{(1, 2, 2)}$, $\mathbf{(1, 2, -2)}$, 
$\mathbf{(1, -2, 2)}$, and $\mathbf{(1, -2, -2)}$ on 
 the $SO(14)$ point from the 
intersections of the other seven-branes.
Therefore, the superpotential in Type I $SO(10)$ model is
\bea 
{ W_{SSM}}&=&
y_{ij}^{UD} Q_i H Q^R_j 
+ y_{ij}^{E\nu}  L_i H L^R_j + 
  \lambda_1^{ijk} \overline{X}_i L_j \overline{L}_k           
+  \lambda_2^{ijk} X_i L^R_j \overline{L}_k^{R}  
+ \lambda_3^{ij} X'_i L^R_j \overline{\Phi}  
\nnb \\ &&
+  \mu' H H' + \mu_{\Phi} \Phi  \overline{\Phi}
+ \lambda_1^{\prime ij} {1\over {M_{\rm Pl}}} HH \phi_i^{-+} \phi_j^{--}
+ \lambda_2^{\prime ij} {1\over {M_{\rm Pl}}} H'H' \phi_i^{++} \phi_j^{+-}
~,~\, 
\label{S0(10)-I-poten}
\eea
where $y_{ij}^{UD}$,  $y_{ij}^{E \nu}$,  $\lambda_1^{ijk}$,
$ \lambda_2^{ijk}$, $\lambda_3^{ij}$, $\lambda_1^{\prime ij}$,
and $\lambda_2^{\prime ij}$
are Yukawa couplings. Similar to the $SU(5)$ model,
the $\mu'$ and $\mu_{\Phi}$ terms 
can be generated via the instanton
effects or Higgs mechanism. Interestingly, we can  
have the Yukawa coupling unification for 
the SM quarks and the Yukawa coupling unification for
the SM leptons and neutrinos
at the GUT scale. Also, after $\overline{X}_i$ and $X_i$ obtain the VEVs,  
three linear combinations of the
six $L_i$ fields and three $\overline{L}_i$ fields will obtain
the vector-like masses, and three linear combinations of
the six $L_i^{R}$ fields and three $\overline{L}_i^{R}$ fields will obtain
the vector-like masses. 
 And then these vector-like particles  will be decouple
around the GUT scale.
Thus, only three linear
combinations of six $L_i$ fields  and three linear
combinations of  six $L^R_i$ fields
 will be massless below the GUT scale.
In addition, after $\Phi$ and $\overline{\Phi} $ obtain 
 VEVs, the right-handed neutrinos and some linear 
combinations of $X'_k$ will obtain the Dirac masses.
Thus, with three or more pairs of vector-like particles
$X'_k$ and $\overline{X}'_k$, we can
 generate the active neutrino masses via 
the double seesaw mechanism~\cite{extended}.
In short, although the SM fermion Yukawa matrices
are rank one, the SM fermion
masses and mixings can be generated in the presence
of $H$-fluxes~\cite{Cecotti:2009zf}.
Furthermore, with small $\mu'$ term, we can make sure that
the bidoublet Higgs field $H'$ obtains mass at the high scale
 while  the bidoublet Higgs field $H$
remains light around the electroweak scale by choosing 
suitable VEVs for the SM singlet Higgs fields 
$\phi^{++}_i$, $\phi^{+-}_i$, $\phi^{-+}_i$,
and $\phi^{--}_i$. Then the bidoublet Higgs field
$H'$ can be decoupled, and the $\mu$ term for 
the bidoublet Higgs field $H$ can be around the TeV scale.
Therefore, we can only have the supersymmetric Standard Model
at the low energy.
In particular, to explain the SM gauge couplings
at the string scale, we can consider 
the heavy bidoublet Higgs field $H'$ as the heavy threshold
corrections to the RGE running of the SM gauge couplings. 
The detailed discussions will be given elsewhere.


\subsection{Type II $SO(10)$ Model}


\begin{table}[htb]
\begin{center}
\begin{tabular}{|c|c|c|c|c|c|}
\hline
${\rm Particles }$ & ${\rm Curve}$ & ${\rm Class}$ & $g_{\Sigma}$ &
$L_{\Sigma}$ & $L_{\Sigma}^{\prime n}$\\\hline
$ \left(H,~H'\right)$ & $\Sigma_{H}\text{ (pinched)}$ & $3H-E_{1}-E_{2}$ & $1$ &
$\mathcal{O}_{\Sigma_{H}}(p_1-p_2)^{1/2}$ & $\mathcal{O}_{\Sigma_{H}}$\\\hline
$ \left(\Phi,~\overline{\Phi}\right)$ 
& $\Sigma_{\Phi}\text{ (pinched)}$ & $3H-E_{1}-E_{2}-E_8$ & $1$ &
$\mathcal{O}_{\Sigma_{\Phi}}(p'_1-p'_2)^{1/2}$ & 
$\mathcal{O}_{\Sigma_{\Phi}}(p'_1-p'_2)^{3/2}$\\\hline
$\left(4L^R_{i},~3 Q_j,~2Q^R_k, L_5\right)$ &
$\Sigma_{Q1}$ & $2H-E_1-E_{3}$ & 0 &
$\mathcal{O}_{\Sigma_{Q1}}(1)^{1/2}$ &   $\mathcal{O}_{\Sigma_{Q1}}
(-1)^{5/2}$\\\hline
$\left(4 L_{i},~3Q^R_j,~2 Q_k,~ L_5^{R}\right)$ &
$\Sigma_{QR1}$ & $2H-E_2-E_{3}$ & 0 &
$\mathcal{O}_{\Sigma_{QR1}}(-1)^{1/2}$ &   $\mathcal{O}_{\Sigma_{QR1}}
(-1)^{5/2}$\\\hline
$\left(2\overline{L}_{l},~ \overline{Q}^{R}_1,~ L^R_6\right)$ &
$\Sigma_{LR6}$ & $2H-E_1-E_{4}$ & 0 &
$\mathcal{O}_{\Sigma_{LR6}}(1)^{1/2}$ &   $\mathcal{O}_{\Sigma_{LR6}}
(1)^{1/2}$\\\hline
$\left(2 \overline{L}^{R}_{l},~\overline{Q}_1,~ L_6\right)$ &
$\Sigma_{L6}$ & $2H-E_2-E_{4}$ & 0 &
$\mathcal{O}_{\Sigma_{L6}}(-1)^{1/2}$ &   $\mathcal{O}_{\Sigma_{L6}}
(1)^{1/2}$\\\hline
$\left(2 \overline{L}_{m},~ \overline{Q}^{R}_2,~  L^R_7\right)$ &
$\Sigma_{LR7}$ & $2H-E_1-E_{5}$ & 0 &
$\mathcal{O}_{\Sigma_{LR7}}(1)^{1/2}$ &   $\mathcal{O}_{\Sigma_{LR7}}
(1)^{1/2}$\\\hline
$\left(2 \overline{L}^{R}_{m},~\overline{Q}_2,~ L_7\right)$ &
$\Sigma_{L7}$ & $2H-E_2-E_{5}$ & 0 &
$\mathcal{O}_{\Sigma_{L7}}(-1)^{1/2}$ &   $\mathcal{O}_{\Sigma_{L7}}
(1)^{1/2}$\\\hline
\end{tabular}
\end{center}
\caption{ The particle curves
and the gauge bundle assignments for each curve in the Type II $SO(10)$ model.
Here, $i=1,~2,~3,~4$, $j=1,~2,~3$, $k=4,~5$, $l=1, ~2$, and $m=3, ~4$.}
\label{SO(10)-II}
\end{table}


In Type II $SO(10)$ model, we present the particle curves with 
homology classes and the gauge bundle assignments for each curve  
in Table~\ref{SO(10)-II}. Note that the Higgs triplets in $\mathbf{10}$
of $SO(10)$ do not have zero modes, we solve the doublet-triplet
splitting problem. Also,
we have seven $L_i$ fields,  seven $L_i^R$ fields,
five $Q_i$ fields, five $Q_i^R$ fields, four $\overline{L}_i$ fields,
four $\overline{L}_i^{R}$ fields, two $\overline{Q}_i$ fields, and two
$\overline{Q}_i^{R}$ fields. Thus, for the net numbers of the
chiral superfields, we have three families of the SM
fermions, {\it i.e.}, three $L_i$ fields, three 
$L_i^R$ fields, three $Q_i$ fields, and three $Q_i^R$ fields.
In addition, we assume that the SM fermion and Higgs
 curves $\Sigma_H$,
$\Sigma_{Q1}$, and $\Sigma_{QR1}$   intersect
at one point in $S$ where the gauge symmetry 
is enhanced to $E_7$, and the SM fermion and Higgs
curves  $\Sigma_{Q1}$ and $\Sigma_{\Phi}$ intersect
at the other point in $S$ where the gauge symmetry 
is enhanced to $E_7$ as well.  On the second $E_7$ point,
we introduce the SM singlet fields
$X'_k$ and $\overline{X}'_k$ respectively
with $SO(10)\times U(1)_a\times U(1)_b$ 
quantum numbers $\mathbf{(1, 0, 2)}$ and $\mathbf{(1, 0, -2)}$
from the intersection of the other seven-branes.
In order to preserve the $U(1)_a\times U(1)_b$ symmetry
for the SM fermion Yukawa couplings,
we choose that in 
the small open neighborhoods of these $E_7$ points,
 the Higgs curve $\Sigma_H$ 
satisfies $t_a=0$, the Higgs curve  $\Sigma_{\Phi}$ 
satisfies $t_a-t_b=0$, 
the SM fermion curves $\Sigma_{Q1}$, $\Sigma_{LR6}$ and
$\Sigma_{LR7}$ satisfy $t_a+t_b=0$, and 
the SM fermion curves $\Sigma_{QR1}$,  $\Sigma_{L6}$,
and $\Sigma_{L7}$ satisfy $t_a-t_b=0$.
For simplicity, we assume that one linear combination
of the four $L_i$ fields, the $L_5$, $L_6$, $L_7$ fields, 
and four $\overline{L}_i$ fields form vector-like
particles, and obtain the vector-like masses 
at the GUT scale via the instanton
effects or the Higgs mechanism due to their couplings to 
the SM singlet Higgs fields with
 $SO(10)\times U(1)_a\times U(1)_b$ 
quantum number $\mathbf{(1, 0, -2)}$
 at the intersections of the other seven-branes.
We assume that 
one linear combination
of the four $L^R_i$ fields,  the $L^R_5$, $L^R_6$, $L^R_7$ fields, and 
four $\overline{L}_i^{R}$ fields
form the vector-like
particles, and obtain the vector-like masses at the GUT scale 
via the instanton
effects or the Higgs mechanism due to their couplings to 
the SM singlet Higgs fields with
 $SO(10)\times U(1)_a\times U(1)_b$ 
quantum number $\mathbf{(1, 0, 2)}$
 at the intersections of the other seven-branes.
We assume that
 two $Q_k$ fields and two $\overline{Q}_k$
fields form vector-like
particles, and obtain the vector-like masses at the GUT scale 
via the instanton
effects or Higgs mechanism. And we assume that 
two $Q^R_k$ fields and 
two $\overline{Q}^{R}_k$ fields form vector-like
particles, and obtain the vector-like masses at the GUT scale 
via the instanton
effects or Higgs mechanism. Here we assume that $i=1,~2,~3,~4$, 
and $k=4,~5$. Thus, we have three families of the SM
left-handed leptons and neutrinos from three linear 
combinations of four $L_i$ fields and three families 
of the SM right-handed quarks 
on the curve $\Sigma_{QR1}$, and have three families of the SM
right-handed leptons and neutrinos from three linear 
combinations of four $L^R_i$ fields and three families 
of the SM left-handed quarks 
on the curve $\Sigma_{Q1}$.

Moreover,  the Higgs curve $\Sigma_H$ 
is pinched at a point, where the gauge symmetry is enhanced to $SO(14)$.
We assume that the Higgs curve $\Sigma_H$ 
satisfies $t_1=0$ in the small open neighborhood of the $SO(14)$ point.
Similar to the above subsection,
we introduce the SM singlet Higgs 
fields $\phi^{++}_i$, $\phi^{+-}_i$, $\phi^{-+}_i$,
and $\phi^{--}_i$ respectively
with $SO(10)\times U(1)_1\times U(1)_2$ 
quantum numbers $\mathbf{(1, 2, 2)}$, $\mathbf{(1, 2, -2)}$, 
$\mathbf{(1, -2, 2)}$, and 
$\mathbf{(1, -2, -2)}$ on the $SO(14)$ point from the 
intersections of the other seven-branes.
Therefore, the superpotential in Type II $SO(10)$ model is
\bea 
{ W_{SSM}}&=&
y_{ij}^{UD} Q_i H Q^R_j 
+ y_{ij}^{E\nu}  L_i H L^R_j 
+ \lambda^{ij} X'_i L^R_j \overline{\Phi}  
+  \mu' H H' + \mu_{\Phi} \Phi \overline{\Phi}
\nnb \\ &&
+ \lambda_1^{\prime ij} {1\over {M_{\rm Pl}}} HH \phi_i^{-+} \phi_j^{--}
+ \lambda_2^{\prime ij} {1\over {M_{\rm Pl}}} H'H' \phi_i^{++} \phi_j^{+-}
~,~\, 
\label{S0(10)-II-poten}
\eea
where $y_{ij}^{UD}$,  $y_{ij}^{E \nu}$, $\lambda_3^{ij}$, 
$\lambda_1^{\prime ij}$,
and $\lambda_2^{\prime ij}$
are Yukawa couplings. $\mu'$ and $\mu_{\Phi}$ terms 
can be generated via the instanton
effects or Higgs mechanism. Interestingly, 
we can have the Yukawa coupling unification 
for the SM quarks, and the Yukawa coupling unification 
for the SM leptons and neutrinos
at the GUT scale. Also, the SM fermion
masses and mixings can be generated in the presence
of $H$-fluxes~\cite{Cecotti:2009zf}.
In addition, after $\Phi$ and $\overline{\Phi} $ obtain 
 VEVs, the right-handed neutrinos and some linear 
combinations of $X'_k$ will obtain the Dirac masses.
Thus, with three or more pairs  of vector-like particles
$X'_k$ and $\overline{X}'_k$, we can explain
the neutrino masses and mixings via the double seesaw
mechanism~\cite{extended}.
Similar to the above subsection, we can decouple the bidoublet
Higgs field $H'$ at the high scale
while keep $H$ around the TeV scale.
In short,  at the low energy we obtain the supersymmetric 
Standard Model as well.
In particular, to explain the SM gauge couplings
at the string scale, we can consider 
the heavy bidoublet Higgs field $H'$ as the heavy threshold
corrections to the RGE running of the SM gauge couplings. 
The detailed discussions will be given elsewhere.









\section{Conclusions}

In this paper, we first briefly reviewed the F-theory model 
building. We constructed one $SU(5)$ model and two $SO(10)$ 
models from F-theory.  The $SU(5)$ gauge symmetry can be 
broken down to the SM gauge symmetry by turning on 
the $U(1)_Y$ flux, and the $SO(10)$ gauge symmetry 
can be broken down to the 
$SU(3)_C\times SU(2)_L\times SU(2)_R\times U(1)_{B-L}$
gauge symmetry by turning on the $U(1)_{B-L}$ flux. 
To preserve the $U(1)_a\times U(1)_b$ gauge or global 
symmetries at the triple intersections of the SM fermion and 
Higgs curves, we put the left-handed quarks and
the right-handed quarks on the different matter curves,
and put the left-handed leptons and neutrinos and
the right-handed leptons and neutrinos on the
different matter curves. Thus, in our models,
all the SM fermion Yukawa couplings 
are indeed invariant under the enhanced $U(1)_a \times U(1)_b$ 
gauge or global symmetries at the triple intersections of 
the SM fermion and Higgs curves.  And then our models
can be understood very well from the local field theory
description. Although the SM fermion Yukawa matrices
are rank one, the SM fermion
masses and mixings can be generated in the presence
of $H$-fluxes.
In addition, we can solve the doublet-triplet splitting
problem naturally.  The extra vector-like 
particles can obtain the heavy masses via the instanton effects
or Higgs mechanism and then decouple at
the high scale. Therefore,
we only have the supersymmetric Standard Model 
at the low energy.
Similar to the $SU(5)$ models,
we showed that the SM gauge couplings are also splitted
in our $SO(10)$ models due to the $U(1)_{B-L}$ flux.
Interestingly, we can explain 
the SM gauge couplings at the string scale by
considering the heavy threshold corrections from the extra 
vector-like particles in our $SU(5)$ and $SO(10)$
models.  Furthermore, in the $SU(5)$ model, 
we can have the Yukawa coupling unification for the bottom 
quark and tau lepton. In the $SO(10)$ models, we can
have the Yukawa coupling unification for the top and 
bottom quarks, and the Yukawa coupling unification 
for the tau lepton and tau neutrino.

\begin{acknowledgments}

This research was supported in part 
by the Cambridge-Mitchell Collaboration in Theoretical Cosmology,
and by the Natural Science Foundation of China under grant No. 10821504.

\end{acknowledgments}

\appendix

\section{Breifly Review of del Pezzo Surfaces}

The del Pezzo surfaces $dP_n$, where $n=1,~2,~...,~8$, are
defined by blowing up $n$ generic points of 
$\mathbb{P}^{1}\times\mathbb{P}^{1}$ or 
$\mathbb{P}^2$. The homological group
$H_2(dP_n, Z)$ has the generators
\begin{equation}
H,~E_1, ~E_2,~...,~E_n~,~\,
\end{equation}
where $H$ is the hyperplane class for $P^2$, and $E_i$ are the
exceptional divisors at the blowing up points and are
 isomorphic to $\mathbb{P}^{1}$.
 The intersecting numbers of the generators are
\begin{equation}
H\cdot H=1~,\;\;\:E_{i}\cdot E_{j}=-\delta_{ij}~,\;\;\;H\cdot E_{i}=0~.~\,
\end{equation}
The canonical bundle on $dP_{n}$ is given by
\begin{equation}
K_{dP_{n}}=-c_{1}(dP_{n})=-3H+\sum_{i=1}^{n}E_{i}.
\end{equation}
For $n\geq3$,  we can define the generators as follows
\begin{equation}
\alpha_i=E_i-E_{i+1}~,~~~{\rm where}~~i=1,~2,...,~n-1~,~\,
\end{equation}
\begin{equation}
\alpha_n=H-E_1-E_2-E_3~.~\,
\end{equation}
Thus, all the generators $\alpha_i$ is perpendicular to the canonical
class $K_{dP_{n}}$. And
 the intersection products are equal to the negative Cartan
matrix of the Lie algebra $E_n$, and can be considered as simple 
roots. 

The  curves $\Sigma_i$ in $dP_n$ where the particles are localized 
 must be divisors of $S$. And the genus for curve $\Sigma_i$ is
given by 
\begin{equation}
2 g_i-2~=~[\Sigma_i]\cdot ([\Sigma_i]+K_{dP_{k}})~.~\,
\end{equation}

For a line bundle $L$ on the surface $dP_{n}$ with
\begin{equation}
c_{1}(L)=\sum_{i=1}^{n}a_{i}E_{i},
\end{equation}
where $a_{i}a_{j}<0$ for some $i\neq j $,  the K\"ahler form
$J_{dP_{n}}$ can be constructed as follows
\cite{Beasley:2008dc}
\begin{equation}
J_{dP_{k}}=b_0H-\sum_{i=1}^{n}b_{i}E_{i},
\end{equation}
where $\sum_{i=1}^k a_{i}b_{i}=0$ and $b_0 \gg b_{i}>0$. By the
construction, it is easy to see that the line bundle $L$ solves
the BPS equation $J_{dP_k}\wedge c_{1}(L)=0$.





\end{document}